\documentstyle[12pt]{article}
\def\be{\begin{equation}}
\def\ee{\end{equation}}
\begin{document}

\hspace{5mm}

\begin{center}

{\bf Band Structure of the Fractional Quantum Hall Effect}

\vspace{1cm}

Gautam Dev and J.K. Jain

\vspace{1cm}

Department of Physics, State University of New York,
Stony Brook, New York, 11790-3800

\end{center}

\vspace{1cm}

The eigenstates of interacting electrons in the fractional quantum Hall
phase typically form fairly well defined  bands in the energy space. We
show that the composite fermion theory gives insight into the
origin of these bands and provides an accurate and complete microscopic
description of the strongly correlated many-body states in the
low-energy  bands.

\pagebreak

As is well known,  the ground state is
separated from the other states by an energy gap at some special filling
factors in the fractional quantum Hall effect (FQHE) [1].
He, Xie, and Zhang (HXZ) [2] have recently pointed out that,
at other filling factors, there is in general a {\em band} of low
energy states which is more or less separated from the other states by a gap.
Since all many-body states are degenerate in the absence of interactions,
the fact that some states split off from the rest implies
non-trivial interaction-induced correlations in these states. We
investigate in this paper the following three questions:
(i) How many states are there in the low energy  band? (ii)
What are the quantum numbers associated
with these states? (iii) What is their  microscopic structure?
We show that the composite fermion (CF) theory of the FQHE [3]
predicts correctly
the number of states in the low energy band as well as the
quantum numbers associated with these states. Most importantly,
the trial wave functions of the CF theory give
a very good {\em microscopic} description of these states.
Furthermore, we find that
the CF theory also provides an accurate description of the first excited
band.

Central to  the CF theory is a mapping between the non-trivial problem
of strongly interacting electrons in the
FQHE regime and the relatively straightforward problem of
weakly interacting electrons in the
integer quantum Hall effect (IQHE) regime [3].
This mapping proceeds in two essential steps: the first provides an intuitive
construction relating the FQHE to the IQHE, and the second makes it
precise in terms of explicit microscopic trial wave functions.
The intuitive construction involves  starting with an (incompressible)
IQHE state with $n$ filled Landau levels (LLs),
attaching an even number ($2p$) of flux quanta
to each electron to convert it into a CF,
and then performing a mean field approximation [4] in which this
flux is smeared to produce an incompressible
FQHE state at   filling factor $\nu=n/(2np+1)$.
This construction suggests the following (unnormalized) ansatz wave
functions for the resulting FQHE state
\be
\chi_{_{n/(2np+1)}}=
\prod_{j<k}(z_{j}-z_{k})^{2p}
\Phi_{n}
\ee
where the subscript denotes the filling factor,
and the position of an electron is
denoted by
$z=x+iy$. The Jastrow factor
attaches $2p$ flux quanta to all electrons in
the state $\Phi_{n}$ to convert them into  CFs.
The Laughlin states [5] are recovered for the special case of  $n=1$.

This construction is generalized  straightforwardly
to arbitrary filling factors as follows [3].
Consider a filling
factor $\nu^{*}$, where  $n < \nu^{*} < n+1$, and assume,
to begin with, non-interacting electrons.
In this case the many-body system contains energy bands separated by
the cyclotron energy $\hbar\omega_{c}$.
The lowest energy band  contains states in which the
lowest $n$ LLs are completely filled and the $(n+1)$th LL is partially
occupied;
the first excited band consists of states
in which one electron has been excited by $\hbar\omega_{c}$; and
so on.
Now, start with these states,
attach an even number ($2p$) of flux quanta to each electron, and
perform the mean field approximation to produce states at
$\nu=\nu^{*}/(2p\nu^{*}+1)$. In this process, the
states within one band
will in general mix with each other,
but it is possible, due to the energy gap, that
those in different bands may not mix. The existence of well defined
bands at low energies in the FQHE state suggests that some of the
low-energy bands are indeed not mixed by the above mean field process.
The CF construction then predicts that the number
of states in these bands at $\nu$ is
equal to that in the corresponding bands
at $\nu^{*}$.

We use the standard spherical geometry [6]
for our numerical calculations,
in which $N$ electrons
move on the surface of a sphere under the influence of a radial
magnetic field ($B$) produced by a magnetic monopole of suitable strength at
the center. Due to the rotational symmetry,
the orbital angular momentum, $L$, and its $L_{3}$ component are good quantum
numbers.
We will label the states by $S$ ($2S=$ integer),
where the total flux through the
surface of the sphere is $2Sh/e$, and the filling factor (in the
thermodynamic limit) is
$\nu=N/2S$.
Guided by the above construction, we write the following
trial wave functions to describe the correspondence between the
states in the lowest energy IQHE and FQHE bands  (generalization to
higher bands is analogous):
\be
\chi_{_{S,\beta}}=
\prod_{j<k}(u_{i}v_{j}-v_{i}u_{j})^{2p}
\Phi_{S^{*},\beta}\;,
\ee
where
the position of an electron is
denoted by the usual spinor components $u$ and $v$ [6];
$S$ and $S^{*}$ are related by
\be
S=S^{*}+p(N-1)\;\;;
\ee
$\chi_{_{S,\beta}}$ are the eigenstates of interacting electrons at $S$;
and $\Phi_{S^{*},\beta}$ are
{\em orthogonal} wave functions
in the lowest energy band of non-interacting electrons at $S^{*}$,
chosen to be eigenstates of $L^{2}$ and $L_{3}$.

The problem thus reduces to determining the low-energy spectrum of
non-interacting electrons at $S^{*}$. The
number of single particle states in the $n$th
LL ($n=1,2,...$) is
\be D^{*}_{n}\equiv 2S^{*}+2n-1\;\;. \ee
When $S^{*}$ corresponds to $n$ filled LLs,
i.e., $S^{*}=S_{n}^{*}\equiv (N-n^{2})/2n$, the lowest-energy band consists
only of one state,
implying incompressibility at
$S= (N-n^{2})/2n+p(N-1)$.
At $S^{*}_{n}<S^{*}
<S_{n+1}^{*}$
the lowest $n$ LLs are filled and the $(n+1)$th LL
contains \be N^{*}_{n+1}\equiv N-n(2S^{*}+n)\ee electrons in
$D^{*}_{n+1}$ states. This implies that there are
\be \frac{D^{*}_{n+1}!}{N^{*}_{n+1}!/;(D^{*}_{n+1}-N^{*}_{n+1})!}\;\; \ee
many-body states in the lowest energy band at $S^{*}$ and, by the CF
construction, also
at $S=S^{*}+p(N-1)$.
Since the angular momentum quantum numbers are identical for  $\Phi$ and
$\chi$ related by Eq.(2),
the CF theory also predicts these quantum numbers
for all the states in the lowest band at $S$.

The number of eigenstates of interacting electrons in the lowest energy band
and their angular momentum quantum numbers, as seen in the
numerical calculations
of HXZ,
are in complete agreement with the
above predictions. Encouraged by this preliminary success, we
proceed to compare the trial wave functions
with the true Coulomb wave functions
for a more direct and rigorous
verification of the CF theory.
Since the numerical calculations are conveniently performed at
$B=\infty$, we need to adiabatically continue our
trial wave functions to this limit. We assume that this can be done
by simply projecting them on to the lowest LL. I.e., we choose as our
$B=\infty$ (unnormalized) trial wave functions
\be
\overline{\chi}_{_{S,\beta}}\equiv {\cal P}\chi_{_{S,\beta}}={\cal P}
\prod_{j<k}(u_{i}v_{j}-v_{i}u_{j})^{2p}
\Phi_{S^{*},\beta}\;,
\label{eq:analogy}
\ee
where ${\cal P}$ is the lowest LL projector.
${\cal P}$ commutes with $L$ [7], and is generally expected to
perturb the wave functions $\chi$ only very gently,
since they are already
predominantly in the lowest LL [3,8]. It has been explicitly shown by Rezayi
and MacDonald [7] for $\nu=2/5$ that
an adiabatic continuation of
$\chi$ to $B=\infty$ essentially produces the projected state
$\overline{\chi}$.

We study a system of six electrons interacting via Coulomb
interactions. Since the energies are independent of
$L_{3}$ for a given $L$ multiplet, it is sufficient to
consider the $L_{3}=0$ sector, with the understanding that one
eigenstate at any $L$ in this sector corresponds to a total of $2L+1$
degenerate eigenstates.
In the following, we will restrict our discussion to the $L_{3}=0$ sector.
In several cases
there is only one state in the lowest energy band for a given $L$  at
$S^{*}$,
so that the trial wave
function for the corresponding
eigenstate at $S$ is uniquely determined, with
no free parameters. In other cases, there are several (up to three in
our work) states at $S^{*}$ with the same $L$.
In these cases, there is some arbitrariness in the choice of the
IQHE states from which the FQHE states are derived.
We choose that {\em orthogonal} set of states at
$S^{*}$ which optimizes the
overlaps. (Details will be given
elsewhere [9].)
Overlaps of the projected trial wave functions
with the
numerical states in the lowest band  are shown in Table I [10].
The states are denoted by $2S$-$L$, and  the parentheses show $K$,
which is the total number of
independent eigenstates at $S$ with angular momentum $L$.

At $2S=15$ ($\nu=1/3$), the lowest energy band contains a single state.
The excited states at $L=2,3,4,5,6$
seem to form a second band (Fig.1), which  is natural to
relate to the
the first excited band at $2S^{*}=5$ ($\nu^{*}=1$).
At first sight, however, there appears to be a serious counting problem:
in addition to the above values of $L$,
the first excited band at $2S^{*}=5$ also contains a state
at $L=1$.
We nevertheless construct trial wave functions $\chi$
and apply ${\cal P}$ to obtain $\overline{\chi}$.
To our pleasant surprise, we find
that {\em the state $\chi$ at $L=1$
is annihilated by the application of ${\cal P}$}.
Thus, quite remarkably, the CF
theory knows that there is no
$L=1$ state in the first excited band at $\nu=1/3$ [11].
The overlaps of the trial states at other values of $L$
with the true Coulomb states
in the excited band
are also shown in Table I.

The large overlaps in all cases
show that the CF theory provides
a good microscopic description of the actual
FQHE states in the low energy bands.
For reference, it
should be kept in mind that a
{\em random} trial wave function with a given $L$
has an overlap of
the order of $1/\sqrt{K}$ with the corresponding true state.
The relatively small overlap for the 15-2 state can be explained due
to the fact that this state is not as well separated from the
other higher energy states at $L=2$
(Fig.1).
It should be noted that a large overlap is in general
much too stringent a
condition for the validity of a trial wave function, and is by no
means necessary. Even
with poor overlap, a trial wave function can be adiabatically
connected to the true physical state and hence contain the relevant
qualitative physics. A large overlap, however, certainly constitutes
a powerful evidence in favor of a trial wave function.

By taking the electrons at $S^{*}$ to be non-interacting,
we have neglected the possibility of a finer structure within the
lowest energy band.
When interactions are turned on, this band will broaden and
may further split into smaller
subbands, possibly translating into
analogous structure at $S$. At some values of $S^{*}$,
the interacting
system can be incompressible even at non-integer fillings,
due to formation of states
related to the above FQHE states by particle-hole symmetry or LL addition,
which will
lead to FQHE at new filling factors via the CF construction [12].
While, in principle, the
subdivision into finer and finer bands can continue {\em ad infinitum},
leading to a hierarchy of states at {\em all}
odd-denominator rational fraction [13],
we do not expect this to be the case for the realistic Coulomb
interaction.

HXZ analyzed their numerical calculations
using the framework of the quasiparticle hierarchy
(QPH) theory of Haldane [6] and Halperin [14].
In this approach, the system at
an intermediate filling is viewed as a collection of
quasiparticles of the ``parent" incompressible state.
HXZ find that the QPH theory is
successful in predicting the number of
states in the lowest energy band as well as their quantum numbers.
However, there
is no direct way of comparing the
pseudo-wavefunctions of the quasiparticles with the electron wave
functions generated in the numerical work.
We also note that while the CF approach provides a uniform
treatment for all filling factors,
in the QPH
scheme, as discovered empirically by HXZ, the quasiholes must be
treated as non-interacting whereas the quasielectrons as interacting (with
a suitable hard-core potential) in order to obtain the right number of
states in the lowest energy band.

Prior to the CF theory, the QHEs at integer and fractional filling
factors were treated as two
essentially unrelated phenomena.
Before closing, we would like to discuss some conceptual issues
concerning the
existence of a fundamental connection between the two.
Even though such a connection
would seem quite natural, and virtually unavoidable from the
phenomenological point of view, the
following questions have been raised with regard to the IQHE-FQHE
mapping of the CF theory:
(i) Why should the higher LLs be invoked in order to understand the FQHE,
a phenomenon that
is believed to occur even in the limit of infinite
$B$?
(ii) How can there be a connection between the IQHE and the FQHE given
that the latter requires interactions in a fundamental manner whereas the
the former can occur even
for non-interacting electrons?
These questions are answered as follows.
(i) There is a large space of
parameters in which the FQHE is observed.  The primary objective of
any theory is to obtain a clear-cut understanding of the phenomenon {\em at
any one point} in this space,  which can then, at least in principle,
be adiabatically continued to the physical point.
No simple microscopic
description of
the FQHE has become possible within the lowest LL,
except in some special cases [5]. The
CF theory asserts that the complete physics of the FQHE becomes
manifest when  a small amount
of hybridization with the higher LLs is allowed (and
an adiabatic continuation
to $B=\infty$ simply requires projection on to the lowest LL).
(ii) The analogy between the IQHE and FQHE does not imply that
interactions are not required for the FQHE.
The crucial role of interactions is to generate CFs.
However, once the CFs are generated, the residual interactions between
them can be neglected (in the simplest cases),
and the FQHE of electrons occurs when
the CFs assume IQHE-like structure to produce incompressibility.

Even though there is no conceptual difficulty with a relationship
between the IQHE and the FQHE, the validity of the IQHE-FQHE
mapping postulated
by the CF theory
(Eq.~\ref{eq:analogy})
can be established only
from an extensive comparison of its consequences with experiments and
numerical studies.
In this paper, we have studied a system of six electrons in the
filling factor range $2/7\leq \nu \leq 2/5$ and found that the
CF theroy
provides a complete and microscopically accurate
account of the entire  low-energy Hilbert space of
interacting electrons at
arbitrary filling factors. Thus, somewhat like in Landau's
fermi liquid theory, there is a one-to-one correspondence
between the low energy
eigenstates of strongly interacting electrons in the FQHE regime and
those of weakly interacting electrons in the IQHE regime.

This work was supported in part by the National Science Foundation
under Grants No. DMR90-20637, PHY89-04035, and the Alfred P. Sloan
Foundation (JKJ).

\pagebreak

\noindent
{\bf References}:

\noindent
[1] See, for a review, {\em The Quantum Hall Effect}, edited by R.E.
Prange and S.M. Girvin (Springer-Verlag, Berlin, 1990), 2nd ed.

\noindent
[2] S. He, X.C. Xie, and F.C. Zhang, Phys. Rev. Lett. {\bf 68}, 3460
(1992).

\noindent
[3] J.K. Jain, Phys. Rev. Lett. {\bf 63}, 199 (1989); Phys. Rev. B{\bf
41}, 7653 (1990); Advances in Phys. {\bf 41}, 105 (1992).

\noindent
[4] R.B. Laughlin, Phys. Rev. Lett. {\bf 60}, 2677 (1988).

\noindent
[5] R.B. Laughlin, Phys. Rev. Lett. {\bf 50}, 1395 (1983).

\noindent
[6] F.D.M. Haldane, Phys. Rev. Lett. {\bf 51}, 605 (1983).

\noindent
[7] E.H. Rezayi and A.H. MacDonald, Phys. Rev. B{\bf 44}, 8395 (1991).

\noindent
[8] N. Trivedi and J.K. Jain, Mod. Phys. Lett. B{\bf 5}, 503 (1991).

\noindent
[9] G. Dev and J.K. Jain, unpublished.

\noindent
[10] Some of the overlaps of Table I have been reported earlier, and are
repeated here for completeness. These are: for 11-0 state in G. Dev and
J.K. Jain, Phys. Rev. B{\bf 45}, 1223 (1992); for 15-0 state in
F.D.M. Haldane and E.H. Rezayi, Phys. Rev. Lett. {\bf 54}, 237 (1985).
Also, the 15-0 and 16-3 trial wave functions are identical to those
proposed by Laughlin [5].

\noindent
[11] The generality of this result for arbitrary number of particles
will be discussed elsewhere. Also, this example shows that sometimes the
projector ${\cal P}$ can affect the states $\chi$ drastically.

\noindent
[12] For example, the interacting system at $2S^{*}=9$
corresponds to $\nu^{*}=2/3$ and is incompressible due to particle-hole
symmetry in the lowest LL. This results in FQHE
at $2S=19$ which corresponds to $\nu=2/7$. We have obtained a trial
wave function for the 2/7 state by diagonalizing the
Coulomb Hamiltonian at $2S^{*}=9$
and multiplying the ground state by
the Jastrow factor. It
has an overlap of 0.9787 with the Coulomb ground state at 2/7.

\noindent
[13] J.K. Jain and V.J. Goldman, Phys. Rev. B{\bf 45}, 1255 (1992).

\noindent
[14] B.I. Halperin, Phys. Rev. Lett. {\bf 52}, 1583 (1984).

\pagebreak

{\bf Table Caption}:

\vspace{.5 cm}

Table I. Overlaps
for all states in the lowest energy band for each
value of $S$ in the range
$11 \leq 2S\leq 19$, and for all states in the first excited band at
$2S=15$. The size of the system is $N=6$ electrons so that the
values $2S=11$, 15, and 19 correspond to $\nu=2/5,$ 1/3,
and 2/7, respectively. The states are denoted by $2S$-$L$
followed by $K$ in parentheses (see the
text for the definition of these symbols).
The states in the first excited band are marked by asterisk.

\vspace{1cm}

{\bf Figure Caption}:

\vspace{.5cm}

Fig. 1  Energy spectrum at $2S=15$. The two lowest energy bands are
delineated by horizontal lines.

\pagebreak

\begin{tabular}{|c|c|c|c|c|c|c|c|c|} \hline
19-0(10)& 19-2(23)&19-3(21)& 19-4(37)& 19-5(32)& 19-6(49)& 19-7(43)&
19-8(56)
&19-9(51) \\
\hline
0.9956&0.9909&0.9945&0.9894&0.9932&0.9894&0.9865&0.9883&
0.9854 \\
0.9694&0.9720&&0.9870&&0.9680&&0.9886& \\
&&&0.9874&&0.9735&&& \\ \hline \hline
19-10(62)&19-12(65)&18-1(13)&18-3(26)&18-4(23)&18-5(34)
&18-6(33)&18-7(43)&18-9(49) \\
\hline
0.9758&0.9754&0.9947&0.9879&0.9886&0.9914&0.9894&0.9822&0.9768
\\
&&&0.9765&&&&& \\ \hline \hline
17-0(8)&17-2(16)&17-4(26)&17-6(34)&
16-3(18)&15-0(6)&15-2(11)*&15-3(9)*&15-4(17)* \\ \hline
0.9714&0.9918&0.9896&0.9815&0.9889&0.9964&0.9484&0.9923&0.9915 \\
\hline \hline
15-5(13)*&15-6(22)*&14-3(14)&13-0(5)&13-2(9)&13-4(14)
&12-1(4)&12-3(7)&11-0(3) \\ \hline
0.9977&0.9800&0.9882&0.9984&0.9841&0.9933&0.9786&0.9956&0.9998 \\
\hline
\end{tabular}

\end{document}